\documentstyle[12pt]{article}
\input psfig
\long\def\comment#1{}
\begin{document}
\title{A Note on Quantum Errors and Their Correction}
\author{Subhash Kak\\
Department of Electrical \& Computer Engineering\\
Louisiana State University\\
Baton Rouge, LA 70803-5901; {\tt kak@ee.lsu.edu}}
\maketitle

\begin{abstract}
This note presents a few
observations on the nonlocal nature of quantum errors and
the expected
performance of the
recently proposed quantum error-correction codes that
are based on the assumption that the errors are
either
bit-flip or phase-flip or both.

\end{abstract}

\section{Introduction}

{\it To disavow an error is to invent retroactively.}\\
\hspace*{2in} ---Johann Wolfgang von Goethe

In a classical information system the basic error is represented by a
$0$ becoming a $1$ or vice versa.
The characterization of such errors is in terms of
an error rate, $\epsilon$,
associated with such flips.
The correction of such errors is achieved by appending 
check bits to a block of information bits.
The redundancy provided by the check bits can be
exploited to determine the location of errors using the
method of syndrome decoding.
These codes are characterized by a certain capacity
for error-correction per block.
Errors at a rate less than the capacity of the
code are {\it completely} corrected.

Now let us look at a quantum system.
Consider a single cell in a quantum register.
The error here can be due to a random unitary
transformation or by entanglement with the environment.
These errors cannot be defined in a graded sense because
of the group property of unitary matrices
and the many different ways in which the entanglements
can be expressed.
Let us consider just the first type of error,
namely that of random unitary transformation.
If the qubit is the state $| 0\rangle$, it can
become 
$a |0\rangle + b | 1 \rangle$.
Likewise, the state $| 00\rangle$ can become
$a | 00\rangle + b | 01 \rangle + c | 10 \rangle + d | 11\rangle $.
In the initialization of the qubit a similar
error can occur\cite{Ka98b}.
If the initialization process consists of collapsing
a random qubit to the basis state $| 0\rangle$, the definition
of the
basis direction can itself have a small error associated
with it. 
This error is analog and so, unlike error in classical
digital systems, it cannot be controlled.
In almost all cases, therefore, the qubits will have
superposition states, although the degree of superposition
may be very low.

From another perspective, classical error-correction codes
map the information bits into codewords in a higher
dimensional space so that if just a few errors occur in the
codeword, their location can, upon decoding, be identified.
This identification is possible because the errors perturb
the codewords, {\it locally}, within small spheres.
Quantum errors, on the other hand, perturb the information
bits, in a {\it nonlocal} sense, to a superposition of many states, so the concept of
controlling all errors by using a higher dimensional
codeword space cannot be directly applied.

According to the positivist understanding of
quantum mechanics, it is essential to speak
from the point of view of the observer
and not ask about any intrinsic information in
a quantum state\cite{Ka98a}.
Let's consider, therefore, the representation of
errors by means of particles in a
register of $N$ states.

We could consider errors to be
equivalent to either $n$ bosons or fermions.
Bosons, in a superpositional state
follow the Bose-Einstein statistics.
The probability of each pattern will
be given by

\begin{equation}
\frac{1}{\left( \begin{array}{c}
                N + n - 1 \\ n
                \end{array}
                  \right) }.
\end{equation}
 
So if there are 3 states and 1 error particle, we can only distinguish
between 3 states: $00,~01~or~10,~11$.
Each of these will have a probability of $\frac{1}{3}$.
To the extent this distribution departs from
that of classical mechanics, it represents nonlocality at
work.

If the particles are fermions, then they are
indistinguishable, 
and with $n$ error objects in $N$ cells, we have
each with the probability

\begin{equation}
\frac{1}{\left( \begin{array}{c}
                N  \\ n
                \end{array}
                  \right) }.
\end{equation}

If states and particles have been identified,
these statistics will be manifested by a group
of particles.
If the cells are isolated then their histories
cannot be described by a single unitary transformation.

Like the particles, the errors will also be
subject to the same statistics.
These statistics imply that the errors will not
be independent, an assumption that is basic to
the error-correction schemes examined in the
literature.

To summarize, important characteristics of quantum errors
that must be considered
are {\it component proliferation},
{\it nonlocal effects} and {\it amplitude error}.
All of these have no parallel in the classical case.
Furthermore,
quantum errors are analog and so the system cannot
be shielded below a given error rate. Such shielding
is possible for 
classical digital systems.

We know that a computation need not require
any expenditure of energy if it is cast in the form
of a reversible process.
A computation which is not reversible must involve
energy dissipation.
Considering conservation of
information+energy to be a fundamental principle, a
correction of random errors in the qubits
by unitary transformations, without any expenditure
of energy, violates this principle.

Can we devise error-correction coding for
quantum systems? To examine this,
consider the problem of protein-folding, believed to 
be NP-complete,
which is, nevertheless, solved efficiently by
Nature.
If a quantum process is at the basis of this
amazing result, then it is almost certain that
reliable or fault-tolerant quantum computing must exist
but, paying heed to the above-mentioned conservation law,
it appears such computing will require some lossy operations.

In this note we examine the currently
investigated models of quantum error-correction
from the point of view of their limitations.
We also consider how quantum errors affect
a computation in comparison with classical errors.

\section{Representing quantum errors}

{\it Sed fugit interea, fugit inreparabile tempus.\\
But meanwhile it is flying, irretrievable time is flying.}\\
\hspace*{2in} ---Virgil

Every unitary matrix can be transformed by a suitable
unitary matrix into a diagonal matrix with all its
elements of unit modulus.
The reverse also being true, quantum errors can play havoc.

The general unitary transformation representing errors
for a qubit is:

\begin{equation}
\frac{1}{\sqrt {||e_1||^2 + ||e_2||^2}} \left[ \begin{array}{cc}
                                  e_1^* & e_2^* \\
                                  e_2 & -e_1 \\
                               \end{array} \right] .
\end{equation}

These errors ultimately change the probabilities of
the qubit being decoded as a $0$ and as a $1$.
From the point of view of the user, when the quantum state
has collapsed to one of its basis states,
it is correct to speak of an error rate.
But such an error rate cannot be directly applied to
the quantum state itself.

Unlike the classical digital case, quantum errors cannot
be completely eliminated because they are essentially analog
in nature.

The unitary matrix (1) represents
an infinite number of cases of error.
The error process is an analog process, 
and so, in general, such errors cannot be corrected.
From the point of view of the qubits, it is a
nonlocal process.

If it is assumed that the error process can be represented
by a small rotation and the initial
state is either a $0$ or a $1$, then this rotation
will generate a superposition of the two states 
but the relative amplitudes will be different
and these could be exploited in some specific situations
to determine the starting state.
But, obviously, such examples represent trivial cases.

The error process may be usefully represented 
by a process of quantum diffusions and phase
rotations.

Shor\cite{Sh95} showed how the decoherence
in a qubit could be corrected by a system
of triple redundancy coding where each qubit is
encoded into nine qubits as follows:

\[|0\rangle \rightarrow \frac{1}{2\sqrt 2}  
( |000\rangle + | 111\rangle )
( |000\rangle + | 111\rangle )
( |000\rangle + | 111\rangle )\],

\begin{equation}
|1\rangle \rightarrow \frac{1}{2\sqrt 2}  
( |000\rangle - | 111\rangle )
( |000\rangle - | 111\rangle )
( |000\rangle - | 111\rangle ).
\end{equation}

Shor considers the decoherence process to be one where
a qubit decays into a weighted amplitude
superposition of its basis states.
In parallel to the assumption of independence of
noise in classical information theory,
Shor
assumes that only one qubit out of the
total of nine decoheres.
Using
a Bell basis, Shor then shows that one can determine the
error and correct it.

But this system does not work if more than one
qubit is in error.
Since quantum error is analog, each qubit will be
in some error and so this scheme will, in practice,
not be useful in {\it completely} eliminating
errors.

The question of decoherence, or error, must be considered as
a function of time. 
One may use the exponential function $\lambda e^{-\lambda t}$
as a measure of the decoherence probability of the
amplitude of the qubit.
The measure of decoherence that has taken place by time $t$
will then be given by the probability, $p_t$:

\begin{equation}
p_t = 1 - \lambda e^{-\lambda t}.
\end{equation}

In other words, by time $t$, the amplitude of the
qubit would have decayed to a fraction
$(1 - \lambda e^{-\lambda t})$ of its original value.
At any time $t$, there is a $100 \%$ chance that the
probability amplitude of the initial state will
be a fraction $\alpha_k < 1$ of the
initial amplitude.

If we consider a rotation error in each qubit through angle $\theta$,
there exists some $\theta_k$ so that the probability

\begin{equation}
Prob ( \theta > \theta_k) \rightarrow 1.
\end{equation}

This means that we cannot represent the qubit error
probability by an assumed value $p$ as was done
by Shor in analogy with the classical case.
In other words, there can be no guarantee of
eliminating decoherence.

\section{Recently proposed error-correction codes}

{\it The fox knows many things---the hedgehog one {\em big} one.}\\
\hspace*{2in} ---Archilochus

The recently proposed models of
quantum error-correction codes assume
that the error in the 
qubit state $a |0\rangle + b | 1 \rangle$
can be either a bit flip $ |0 \rangle \leftrightarrow | 1 \rangle$,
a phase flip between the relative phases of $| 0\rangle$
and $| 1 \rangle $, or both \cite{St96,Sh96,Pr97}.

In other words, the errors are supposed to take the pair 
of amplitudes $(a,b)$
to either $(b,a)$, $(a, -b)$, or $(-b,a)$.

But these three cases represent a vanishingly small subset
of all the random unitary transformations associated
with arbitrary error.
These are just three of the infinity of rotations 
and diffusions that the
qubit can be subject to.
The assumed errors,
which are all local,
do not, therefore, constitute a distinguished set on
any physical basis.

In one proposed error-correction code,
each of
the states
$ |0\rangle $ or $  | 1 \rangle$
is represented by
a 7-qubit code, where the 
strings of the codewords represent the
codewords of the single-error correcting
Hamming code, the details of which we don't
need to get into here.
The code for $| 0 \rangle$ has an even number of
$1$s and the code for $|1 \rangle$ has an odd number
of $1$s.

\[| 0\rangle_{code} = \frac{1}{\sqrt8} (|0000000\rangle + |0001111\rangle+
|0110011\rangle + |0111100\rangle\\\]

\begin{equation}
 + |1010101\rangle
+|1011010\rangle +| 1100110\rangle + |1101001\rangle),
\end{equation}

\[|1\rangle_{code} = \frac{1}{\sqrt8} (|1111111\rangle + |1110000\rangle+
|1001100\rangle + |1000011\rangle \\\]

\begin{equation}
+ |0101010\rangle
+|0100101\rangle +| 0011001\rangle + |0010110\rangle).
\end{equation}

As mentioned before, the errors are assumed to be either in terms of
phase-flips or bit-flips.
Now further ancilla bits--- three in total--- are augmented that compute the
syndrome values.
The bit-flips, so long as limited to one in each group, can
be computed directly from the syndrome.
The phase-flips are likewise computed, but only after a change of
the bases has been performed.

Without going into the details of these steps, which are
a straightforward generalization of classical error
correction theory, it is clear that the assumption
of single phase and bit-flips is very restrictive.

In reality, errors in the 7-qubit words will generate a
superposition state of 128 sequences,
rather than the 16 sequences of equations (5) and (6), together with
16 other sequences of one-bit errors, where the errors
in the amplitudes are limited to the phase-flips mentioned
above.
{\it All kinds of bit-flips}, as well as
modificaitons of the amplitudes will be a part of
the quantum state.

We can represent the state, with the appropriate
phase shifts associated with
each of the 128 component states, as follows:

\begin{equation}
 |\phi\rangle = e^{i \theta_{1} } a_1 |0000000\rangle +
 e^{i \theta_{2} } a_2 |0000001\rangle + . . . +
 e^{i \theta_{N} } a_N |1111111\rangle)
\end{equation}

While the amplitudes of the newly generated components
will be small, they would, nevertheless, have a
non-zero error probability.
These components, cannot be corrected by the code
and will, therefore, contribute to an residual
error probability.

The amplitudes implied by (7) will, for the 16 sequences
of the original codeword
after the error has enlarged the set, be somewhat different from 
the original values.
So if we speak just of the 16 sequences
the amplitudes cannot be preserved without error.

Furthermore, the phase errors in (7) cannot be corrected.
These phases are of crucial importance in 
many recent quantum algorithms.

It is normally understood that in classical systems if
error rate is smaller than a certain value, the error-correction
system will correct it.
In the quantum error-correction systems, this important
criterion is violated.
Only certain specific errors are corrected, others even
if smaller, are not.

In summary,
 the proposed models are based on a
local error model while real errors
are nonlocal where we must consider the issues of
component proliferation and amplitude errors.
These codes are not capable of completely
correcting small errors that cause
new entangled component states to be created.

\section{The sensitivity to errors}
The nonlocal nature of the quantum errors is seen
clearly in the sensitivity characteristics of these
errors.

Consider that some data sets related to a problem are being
simultaneously processed by
a quantum machine.
Assume that by some process of phase switching and diffusion
the amplitude of the desired solution out of the entire set is slowly
increased at the expense of the others.
Nearing the end of the computation, the sensitivity of the
computations to errors will increase dramatically,
because the errors will, proportionately, increase for
the smaller amplitudes.
To see it differently, it will be much harder to reverse the
computation if the change in the amplitude or phase
is proportionally greater.

This means that the
``cost'' of quantum error-correction will depend on the
state of the computing system.
Even in the absence of errors, the sensitivity
will change as the state evolves,
a result, no doubt, due to the nonlocal nature of quantum errors.
 
These errors can be considered to be present 
at the stage of state preparation and through
the continuing interaction with the environment
and also due to the errors in the applied
transformations to the data.
In addition, there may exist nonlocal correlations
of qubits with those in the environment. The
effect of such correlations will be unpredictable.

Quantum errors
cannot be localized. For example,
when speaking of rotation errors, there always exists some
$\theta_k > 0$  so that $prob (\theta > \theta_k) \rightarrow 1$.

When doing numerical calculations on a computer, it is
essential to have an operating regime that provides
reliable, fault-tolerant processing.
Such regimes exist in classical computing.
But the models currently under examiniation for 
quantum computing cannot eliminate
errors completely.

The method of
syndrome decoding, adapted from the
theory of classical error-correcting codes,
appears not to be the answer to the problem of fault-tolerant
quantum computing.
New approaches to error-correction need to be investigated.

\section{Conclusions}
Nonlocality, related both to the evolution of the
quantum information system and errors, 
defines a context in which error-correction based
on syndrome decoding will not work.

How should error-correction be defined then?
Perhaps through a system akin to associative
learning in spin glasses.

\end{document}